\documentclass[conference]{IEEEtran}

\usepackage{cite}
\usepackage{amsmath}
\usepackage{tabularx}
\usepackage{bm}
\usepackage{amssymb}
\usepackage{comment}
\usepackage{array}
\usepackage{mathtools}
\usepackage[straightvoltages]{circuitikz}
\usepackage[titlenumbered]{algorithm2e}
\usepackage{pgfplots}
\pgfplotsset{compat=1.14}
\usepackage{subcaption}
\usepackage{multirow}
\usepackage{mathrsfs}
\usepackage{adjustbox}
\usepackage{graphicx}
\usepackage[english]{babel}
\usepackage{blindtext}
\graphicspath{ {./figures/} }
\def\BibTeX{{\rm B\kern-.05em{\sc i\kern-.025em b}\kern-.08em
    T\kern-.1667em\lower.7ex\hbox{E}\kern-.125emX}}

\pgfplotscreateplotcyclelist{mydefaultlist}{%
  black,mark=star\\%
  blue,every mark/.append style={fill=blue!80!black},mark=otimes\\%
  cyan,every mark/.append style={fill=blue!80!black, solid},mark=otimes\\%
  brown!60!black,densely dashed,every mark/.append style={fill=brown!80!black, solid},mark=triangle\\%
  orange,densely dashed,every mark/.append style={solid},mark=diamond\\%
  green,densely dashed,every mark/.append style={solid,fill=green!80!black, solid},mark=oplus\\%
  red,densely dotted,every mark/.append style={fill=red!80!black, solid},mark=asterisk\\%
}

 \IEEEsettopmargin[]{t}{1in}

\begin{document}
	
	\newcommand{\Transpose}{^{\mathsf{T}}}
	\newcommand{\Hermitian}{^{\mathsf{H}}}
	\newcommand{\E}[1]{\mathrm{E} \left[ #1 \right]}
	\newcommand{\Var}[1]{\mathrm{Var} \left[ #1 \right]}
	\newcommand{\Cov}[1]{\mathrm{Cov} \left[ #1 \right]}
	\newcommand{\norm}[1]{\left\lVert#1\right\rVert}
	\newcommand{\mbeq}{\overset{!}{=}}
	\newcommand{\e}{\textrm{e}}
	\newcommand{\T}[1]{T\left(#1\right)}
	\newcommand{\Tn}[2]{T_{#1}\left(#2\right)}
	\newcommand{\diag}[1]{\mathrm{diag} \left( #1 \right)}
	\newcommand{\Prob}[1]{\mathrm{P}\left(#1\right)}
	\renewcommand*{\mathellipsis}{%
	  \mathinner{{\ldotp}{\ldotp}{\ldotp}}%
	}

	\title{Distributed Joint Multi-cell Optimization of IRS Parameters with Linear Precoders}

	\author{
		\IEEEauthorblockN{R. Wiesmayr$^1$, M. Honig$^2$, M. Joham$^1$, W. Utschick$^1$}
		\IEEEauthorblockA{$^1$Department of ECE, Technical University of Munich, Munich, Germany, \textit{\{reinhard.wiesmayr, joham, utschick\}}@tum.de}
		\IEEEauthorblockA{$^2$Department of ECE, Northwestern University, Evanston, IL, mh@ece.northwestern.edu}
	}

	\maketitle

	\begin{abstract} \label{sec:abstract}
	We present distributed methods for jointly optimizing Intelligent Reflecting Surface (IRS) 
	phase-shifts and beamformers in a cellular network.
	The proposed schemes require knowledge of only the intra-cell training sequences and corresponding received signals
	without explicit channel estimation. Instead, an SINR objective is estimated via sample means
	and maximized directly. This automatically includes and mitigates both intra- and inter-cell interference
	provided that the uplink training is synchronized across cells. Different schemes are considered
	that limit the set of known training sequences from interferers.
	With MIMO links an iterative synchronous bi-directional training scheme jointly optimizes the IRS parameters
	with the beamformers and combiners. Simulation results show that the proposed distributed methods
	show a modest performance degradation compared to centralized channel estimation schemes, 
	which estimate and exchange all cross-channels between cells, and perform significantly better 
	than channel estimation schemes which ignore the inter-cell interference.
	\end{abstract}


	\section{Introduction} \label{sec:introduction}
		Intelligent reflecting surfaces (IRSs) have been proposed as a key technology in the evolution from 5G to 6G networks.
		By controlling the phase-shift and attenuation of the reflected electromagnetic wave, an IRS can potentially
		improve channel conditions in the mid-band range, and overcome blocked Line of Sight (LoS) conditions 
		and increase the channel rank in mmWave bands \cite{Gong2020, Wu2019}.
		In both cases, optimizing IRS phase shifts according to a particular performance metric, such as sum rate, requires 
		knowledge of Channel State Information (CSI). 	
		A challenge is how to optimize and adapt the IRS phase-shift parameters jointly with beamformers 
		while minimizing both training overhead and computational complexity.
		
		Channel estimation schemes for IRS assisted MIMO channels have been studied in several papers,
		e.g., \cite{Zhou2021a,Jensen2020,Liu2020,Jiang2021,Taha2020}. That work has focused on a single cell,
		and estimates all composite channels containing the IRS given the set of users. The training overhead generally scales 
		with the number of IRS elements.
		For multi-cell (MC) systems, channel estimation requires coordination among cells,
		since all interferers' pilot sequences in surrounding cells must be known for estimating their cross-channels.
		Further coordination is needed to avoid pilot contamination.
		A joint optimization that maximizes a global performance metric such as sum rate across the cells
		must be centralized, with all CSI (including all direct paths) collected at a single location.
		The estimation and coordination overhead therefore becomes excessive as the size of the network increases.
		
		We propose an alternative distributed method for jointly optimizing IRS parameters with linear precoders
		and combiners that  does not rely on direct channel estimation. Rather, the IRS parameters and downlink (DL) beamformers
		are optimized directly to maximize an estimated Signal-to-Interference-plus-Noise (SINR) criterion,
		where the estimate is obtained at each Base Terminal Station (BTS) from local uplink (UL) training and received signals 
		(assuming uplink/downlink (UL/DL) reciprocity). This implicitly depends on all CSI, but does not require the level 
		of coordination required for collecting all channel estimates in MC systems. 
		Only synchronization of UL/DL pilots is needed. 
		In contrast to previous work on IRS-enhanced cellular systems (e.g., \cite{Pan2020}),
		we consider a MC scenario with an IRS in each cell that can
		simultaneously increase received power to users within the cell and help to suppress all
		sources of interference. The algorithm jointly adapts the IRS parameters in each cell
		with the beamformers assuming knowledge of only those training sequences for the users
		being served within the cell.

		We present numerical results for MISO and MIMO channels. With MISO channels only one round
		of UL training is needed to estimate all filter and IRS parameters. With MIMO
		channels we extend the distributed bi-directional training method in \cite{Shi2014} to optimize
		the beamformers and combiners jointly with the IRS parameters. Specifically, synchronous 
		UL training jointly optimizes the IRS parameters and beamformers for the DL,
		and synchronous DL training optimizes the combiners at the receivers, which are
		used as beamformers for UL training.
		As in \cite{Zhou2021a} and \cite{Jensen2020}, we send these pilots for each fixed set of IRS elements
		in an orthogonal basis set. The optimization then determines the combining coefficients,
		and the IRS phase shifts are relayed from the BTS to the IRS.
				
		Numerical results compare the sum-rate performance of proposed schemes with
		schemes that estimate CSI. For the examples shown, the proposed distributed scheme in which the BTS
		knows only the training sequences for the intra-cell users performs significantly
		better than CSI schemes which neglect inter-cell interference. 
		Compared to estimating and exchanging all cross-channels, our direct optimization method performs
		similarly if training sequences for all users are known whereas the distributed scheme shows 
		a modest performance degradation.
		This holds for both MISO and MIMO links, although bi-directional training for MIMO links requires somewhat 
		more training than for centralized channel estimation.

	\section{System Model} \label{sec:system_model}
	
\begin{figure}[h]
				\includegraphics[width=0.49\textwidth]{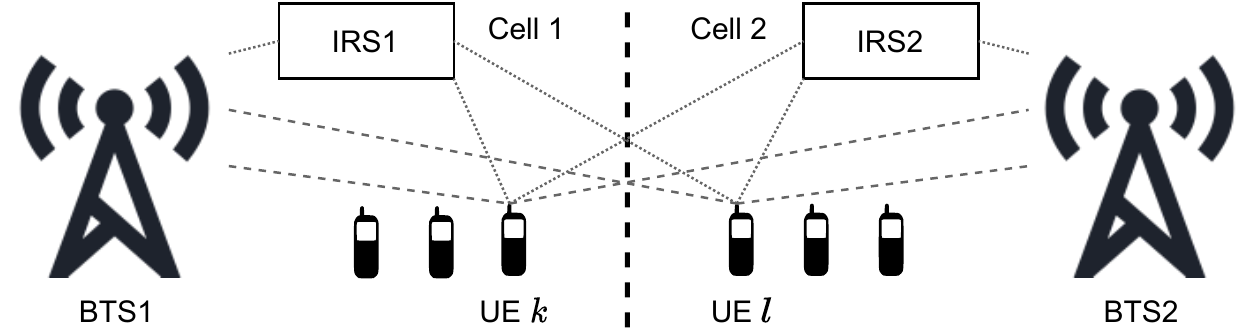}
				\caption{Two-cell model with an IRS in each cell.}
				\label{fig:multicell_system_model}
			\end{figure}

	Figure \ref{fig:multicell_system_model}  illustrates a scenario with two cells with an IRS in each cell.
	Each UE's signal is received at and reflected by each IRS. The proposed methods
	can easily accommodate multiple IRS's within each cell.
	However, we assume each IRS is placed such that 
	there are no reflections from an IRS in one cell to the other IRS and to other cells BTS. This is motivated by the scenario
	in which each IRS has a line-of-sight to its base station and is oriented to point within its cell.

		Following prior work on modeling the effects of a reflective IRS inserted in a MIMO channel \cite{Oezdogan2019}, 
		we write the channel as a product of the incident channel matrix from the transmitter to the IRS, $\bm{H}_{\mathrm{IT}}$, 
		the IRS reflection matrix $\bm{\Theta}= \diag{ \e^{j \phi_1}, \dots, \e^{j\phi_{N_{\mathrm{IRS}}}} }$, 
		and the departing channel matrix from the IRS to the receiver $\bm{H}_{\mathrm{IR}}\Transpose$ 
		(transpose due to channel reciprocity).
		As in prior work (e.g., \cite{Zhang2019,Pan2020}), for purposes of implementation
		we only optimize the phase shifts and do not modify the IRS attenuation.
		
		The system model for the Multi-User (MU) MIMO UL channel with $K$ intra-cell users 
		and $L$ other-cell interferers is then given by
		\begin{equation} \label{eq:mu_mimo_system_model}
			\bm{y} = \sum_{k=1}^K \tilde{\bm{H}}_k \bm{g}_k x_k + \sum_{l=1}^L \tilde{\bm{H}}_l \bm{g}_l x_l + \bm{n} \text{,}
		\end{equation}
		with $\tilde{\bm{H}}_{m}={\bm{H}_{\mathrm{RT}}}_{m}+\bm{H}_{\mathrm{IR}}\Transpose \bm{\Theta} {\bm{H}_{\mathrm{IT}}}_{m} \in \mathbb{C}^{M_{\mathrm{R}}\times N_{\mathrm{T}}}$ for both intra- and other-cell users, 
		UL transmit beamforming vector $\bm{g}_{m}$,
		and $\bm{n}$ is complex Gaussian noise, assumed to be uncorrelated. We will also
		assume that the symbols for all users have variance equal to the transmit power, i.e., $E[|x_m|^2]=\sigma_m^2$.
		
		With linear receivers, the symbol estimate for UE $k$ is $\hat{y}_k=\bm{v}\Hermitian_k \bm{y}$.
		We assume a fixed transmit power allocation and one single stream per UE for simplicity.
		Joint power adaptation with bi-directional training, which can be used to
		determine the number of streams per link, 
		is considered in \cite{Zhou2018}.

		We will consider joint optimization of the IRS parameters $\bm{\Theta}$ with $\bm{v}_k$ on the UL,
		which is then used for the DL beamformer.
		The UL model is then for uplink training, given the UL precoders $\bm{g}_k$'s.
		In the MISO scenario, those UL precoders are scalars, normalized to one, whereas in the MIMO scenario
		the uplink training, which jointly estimates $\bm{\Theta}$ and $\bm{v}_k$, must iterate with DL training 
		to optimize the combiners $\bm{g}_k$.

		We will start with MISO DL channels, where the proposed method consists of a single backward step for
		training and estimation.
		The corresponding MU SIMO UL channel with $K$ intra-cell users and $L$ other-cell interferers is then
			\begin{equation} \label{eq:mu_simo_system_model}
				\bm{y} = \sum_{k=1}^K \tilde{\bm{h}}_k x_k + \sum_{l=1}^L \tilde{\bm{h}}_l x_l + \bm{n} \text{,}
			\end{equation}
			with the IRS-assisted channels $\tilde{\bm{h}}_{k}={\bm{h}_{\mathrm{RT}}}_{k}+\bm{H}_{\mathrm{IR}}\Transpose \bm{\Theta} {\bm{h}_{\mathrm{IT}}}_{k}$
			where $\bm{h}_{\mathrm{RT}}$ is the direct channel and the other product is the IRS product channel.
			The BTS applies the linear receive filter $\bm{v}_k$ to obtain the estimate $\hat{y}_k = \bm{v}_k\Hermitian \bm{y}$ with
			\begin{multline} \label{eq:sinr_mu_simo_ul}
				\mathrm{SINR}_k = \\
				\frac{\left|\bm{v}_k\Hermitian \tilde{\bm{h}}_k \right|^2 \sigma_k^2 }{\sum_{\substack{k^{\prime}=1 \\ k^{\prime} \neq k}}^K \left|\bm{v}_k\Hermitian \tilde{\bm{h}}_{k^\prime} \right|^2 \sigma_{k^\prime}^2  + \sum_{l=1}^L \left|\bm{v}_k\Hermitian \tilde{\bm{h}}_{l} \right|^2 \sigma_l^2  +  \sigma_n^2 \bm{v}_k\Hermitian  \bm{v}_k} \textit{,}
			\end{multline}
			where $\sigma_n^2$ is the noise variance.
			Since we assume linear precoders and combiners, our performance objective is the approximate sum rate:
			\begin{equation} \label{eq:sum_rate_ul_mu_simo}
				\max_{\bm{V}=  \{\bm{v}_1 \cdots \bm{v}_K \},\bm{\Theta}} \sum_{k=1}^K \log_2\left(1+\mathrm{SINR}_k\right)\text{.}
			\end{equation}
			which holds if the interference is Gaussian.
			
			For the CSI-based benchmark, we optimize \eqref{eq:sum_rate_ul_mu_simo} given estimated channels from training.
			The maximizing linear filter $\bm{V}$ in \eqref{eq:sum_rate_ul_mu_simo} is then the MMSE filter, a function of $\bm{\Theta}$ and the channel estimates.
			We will use bi-directional (alternating) optimization (Max-SINR algorithm) \cite{Shi2014}, 
			taking the scaled UL receive filters 
			as DL precoders. This is motivated by UL-DL reciprocity with linear beamformers/combiners,
			and has been observed to be near-optimal for cases of interest.
			The interference suppression properties of the optimized beamformers and combiners
			will help to mitigate signal power transmission to the $L$ interfering UEs of adjacent cells.
			
			In contrast to the CSI-based schemes, our proposed scheme estimates the cost-function
			\eqref{eq:sum_rate_ul_mu_simo} directly, based on training observations.
			For the DL, each UE receives signals from all BTSs via its direct channel and also from each IRS
			since UL-DL channel reciprocity holds from the UE's perspective. 
			This setup allows IRS optimization in the UL direction independently from other cells, 
			since each IRS only effects UL channels to its connected BTS.
			Thus, each BTS can optimize $\bm{V}$ and $\bm{\Theta}$ independently by maximizing \eqref{eq:sum_rate_ul_mu_simo}.

	\section{Direct IRS Optimization} \label{sec:sample_based_irs_optimization_method}
		Our approach is to express the objective \eqref{eq:sum_rate_ul_mu_simo} in terms of correlations
		containing the pilots, received signal, and IRS phase shifts, which we can estimate directly.
		We adopt a similar training protocol, as for the channel estimation schemes in \cite{Zhou2021a, Jensen2020}:
		To optimize all $N_{\mathrm{IRS}}$ IRS phases, and distinguish the direct from reflected channels, 
		all UEs synchronously transmit their predefined 
		uplink training sequences $\{\bm{b}_k\}$ with
		IRS phase shifts set to predefined values $\bm{\Theta}_j$, $j=1,\cdots, N_{\mathrm{IRS}}+1$.
		That is, the training sequences are repeated $N_{\mathrm{IRS}}+1$ times,
		where the sequence of $\bm{\Theta}_j$'s are taken from a codebook
		$\mathcal{D}=\left\{\bm{\Theta}_1,\dots,\bm{\Theta}_{N_{\mathrm{IRS}}+1}\right\}$,
		consisting of diagonal orthogonal matrices $\bm{\Theta}_j$. This takes into account
		the implicit estimation of the direct path.
		
		We start by assuming that $\bm{\Theta}_j = \bm{\tilde{\Theta}}_j$, which is diagonal where the diagonal components
		form the $j$th unit vector (all zeros except the $j$th component, which is one),
		and define $\bm{y}_{\bm{\tilde{\Theta}}_j}$ as the corresponding received signal.
		We will refer to this set $\tilde{\mathcal{D}}= \{ \bm{\tilde{\Theta}}_1 ,\cdots, \bm{\tilde{\Theta}}_{N_\mathrm{IRS}},\bm{0}\}$ as the {\em canonical} set.
		To determine the IRS phases, we express $\bm{\Theta} = \sum_{j=1}^{N_{\mathrm{IRS}}} w_j \bm{\tilde{\Theta}}_j$
		and optimize the objective over the phases of the complex unit-norm combining coefficients $w_j$.
		The weight vector $\bm{w} = \{w_1 , \cdots , w_{N_{\mathrm{IRS}}} \}$ then contains the IRS phase shifts,
		to be optimized.
		Also, $\bm{y}_{\bm{0}}$ is the received signal for the direct path only, i.e., $\bm{\Theta} = \bm{0}$.

		Assuming the pilots symbols from different users, $b_k$ and $b_m$, $k\neq m$, are uncorrelated, we can write
		\begin{multline} \label{eq:canonical_ms_mimo_terms}
      \E{\left[\sum_{j=1}^{N_{\mathrm{IRS}}} \left(\bm{y}_{\bm{\tilde{\Theta}}_j}-\bm{y}_{\bm{0}}\right)w_j + \bm{y}_{\bm{0}}\right]b_k^*}\\
       = \left({\bm{h}_{\mathrm{RT}}}_k+\bm{H}_{\mathrm{IR}}\Transpose \diag{\bm{w}} {\bm{h}_{\mathrm{IT}}}_k\right)\sigma_k^2
       = \tilde{\bm{h}}_k \sigma_k^2 \text{.}
    \end{multline}
    where $\sigma_k$ is a scale factor for the binary pilots needed to satisfy the power constraint.
	Hence we can estimate $\tilde{\bm{h}}_k$ directly in terms of the correlations on the left-hand side. 
	Substituting those estimates in \eqref{eq:sinr_mu_simo_ul}, and given an estimate for $\sigma_n^2$, we can
	then maximize the SINR over the $\bm{w}$ and $\bm{v}$'s.
	
	Rather than isolating each IRS element sequentially, we can instead activate all IRS elements,
	and take the entries in $\mathcal{D}$ to be unitary. That is, let $\bm{A}$ be an 
	unitary (e.g., DFT or Hadamard) matrix with columns $\bm{a}_j$.
    Analogue to \cite{Zhou2021a}, we construct the diagonal IRS training matrix $\bm{\Theta}_j$ with
    diagonal components 
    \begin{equation}
    \bm{\theta}_j = \mathrm{diag}\left(\bm{\Theta}_j\right) = \bar{\bm{a}}_j=[a_{j,2},\dots,a_{j,N_{\mathrm{IRS}}+1}]\Hermitian .
    \end{equation}
	We then reconstruct $\bm{y}_{\bm{\tilde{\Theta}}_j}$, the received signal 
	for the $j$th element in the canonical set $\tilde{\mathcal{D}}$ as follows.
	By activating all elements of the IRS, this scheme maximizes the received SINR while
	allowing direct estimation of $\tilde{\bm{h}}_k$.
    
		Let
		$\bm{Y}_{\bm{w}}=[\bm{y}_{\bm{w}}(1),\dots,\bm{y}_{\bm{w}}(T)]$ denote the received signals
		with IRS phase shifts $|\bm{w}|=\bm{1}$.
		\begin{enumerate}
			\item All $K+L$ users synchronously send their training sequence $\bm{b}_k\Hermitian = \{b_k(1),\dots,b_k(T)\}$ $N_{\mathrm{IRS}}+1$ times, with $\mathrm{diag}\left(\bm{\Theta}_j\right)=\bar{\bm{a}}_j$, $j = 1,\dots,N_{\mathrm{IRS}}+1$.
			\item Calculate the received signals for the direct channel and with canonical $\tilde{\bm{\theta}}_j = \diag{\tilde{\bm{\Theta}}_j}$ as
			\begin{align}
				\bm{y}_{\bm{0}}(t) & =  \frac{1}{N_{\mathrm{IRS}}+1}\left[\bm{y}_{\bm{\Theta}_1}(t), \dots, \bm{y}_{{\bm{\Theta}}_{N_{\mathrm{IRS}}+1}}(t)\right]\bm{a}_1 \text{,}\\
				\nonumber
				\bm{y}_{\tilde{\bm{\Theta}}_j}(t) & = 
				\frac{1}{N_{\mathrm{IRS}}+1}\cdot\\
				& \left[\bm{y}_{\bm{\Theta}_1}(t), \dots, \bm{y}_{\bm{\Theta}_{N_{\mathrm{IRS}}+1}}(t)\right]\bm{a}_{j+1} + \bm{y}_{\bm{0}}(t) \text{.}
			\end{align}
			\item Taking into account the still to be determined phase-shifts $\bm{w}$, we calculate $\bm{Y}_{\bm{w}}$ as in \eqref{eq:canonical_ms_mimo_terms}, i.e.,
		\end{enumerate}
		\begin{equation}
			\bm{y}_{\bm{w}}(t)=\left(\left[\bm{y}_{\tilde{\bm{\Theta}}_1}(t), \dots, \bm{y}_{\tilde{\bm{\Theta}}_{N_{\mathrm{IRS}}}}(t)\right]- \bm{y_{\bm{0}}}(t)\bm{1}\Transpose_{N_{\mathrm{IRS}}} \right)\bm{w} + \bm{y}_{\bm{0}}(t) \text{.}
		\end{equation} 
		
		To optimize the beamformers jointly with IRS phase shifts
		we minimize the Least Squares (LS) objective
		$\norm{\bm{b}_k\Hermitian-\bm{v}_k\Hermitian\bm{Y}_{\bm{w}}}_F^2$, i.e.,
		\begin{equation} \label{eq:ls_filter_v_samples}
			\bm{v}_k = \left(\bm{Y}_{\bm{w}} \bm{Y}_{\bm{w}}\Hermitian\right)^{-1}\bm{Y}_{\bm{w}} \bm{b}_k\text{.}
		\end{equation}
		The LS filter implicitly accounts for the $L$ other-cell interferers in $\bm{Y}_{\bm{w}}$, and converges 
		to the MMSE filter as $T\to\infty$.

		We now estimate the correlations in the SINR objective as sample averages.
		Specifically, recall that $\E{\bm{v}_k\Hermitian\bm{y}_{\bm{w}} b^*_k} = \bm{v}_k\left({\bm{h}_{\mathrm{RT}}}_k+\bm{H}_{\mathrm{IR}}\Transpose \diag{\bm{w}} {\bm{h}_{\mathrm{IT}}}_k\right)\sigma_k^2$, 
		where we can obtain an unbiased and asymptotically efficient estimate of the correlation as
		\begin{equation} \label{eq:correlation_sample_wise}
			\frac{1}{T}\bm{v}_k\Hermitian(\bm{w}) \bm{Y}_{\bm{w}}\bm{b}_k \xrightarrow{T \to\infty}
			\E{\bm{v}_k\Hermitian\bm{y}_{\bm{w}} b^*_k}\text{.}
		\end{equation}
		When substituting for the LS filter $\bm{v}_k$ from \eqref{eq:ls_filter_v_samples}, \eqref{eq:correlation_sample_wise} can be rewritten as $\frac{1}{T}\bm{b}_k\Hermitian\mathbf{P}_{\bm{Y}_{\bm{w}}\Hermitian}\bm{b}_k$, where $\mathbf{P}_{\bm{Y}_{\bm{w}}\Hermitian} = \bm{Y}_{\bm{w}}\Hermitian\left(\bm{Y}_{\bm{w}} \bm{Y}_{\bm{w}}\Hermitian\right)^{-1} \bm{Y}_{\bm{w}}$ is the orthogonal projection operator onto the column-space of $\bm{Y}_{\bm{w}}\Hermitian$.
		The interference terms in \eqref{eq:sinr_mu_simo_ul} can be estimated by the cross-correlation with the training sequences of intra- or inter-cell UEs, i.e., $\bm{b}_k\Hermitian\mathbf{P}_{\bm{Y}_{\bm{w}}\Hermitian}\bm{b}_{k^\prime}$, or $\bm{b}_k\Hermitian\mathbf{P}_{\bm{Y}_{\bm{w}}\Hermitian}\bm{b}_l$.

		Similarly, the inner products in \eqref{eq:sinr_mu_simo_ul} are estimated by training observations, 
		for example, the estimated signal power is
		\begin{equation}
			\left|\bm{v}_k\Hermitian \tilde{\bm{h}}_k \right|^2 \sigma_k^2 \sim \frac{1}{T^2\sigma_k^2} \left|\bm{b}_k\Hermitian\mathbf{P}_{\bm{Y}_{\bm{w}}\Hermitian}\bm{b}_k\right|^2 \text{.}
		\end{equation}
		To complete the direct optimization scheme, we can estimate the background noise variance
		$\sigma_n^2$ from the RF receiver noise figure and temperature, or by blanking a set of transmitters.
		Numerical results show that a coarse estimate suffices, i.e., within a range of 
		$\pm 10\mathrm{dB}$ only showed a negligible performance degradation.
		Also, we present distributed schemes, which do not require an estimate of $\sigma_n^2$.

		We then optimize the estimated sum rate objective over $\bm{w}$.

		\subsection{Distributed Joint IRS and Precoder Optimization} \label{sec:joint_irs_param_precoding_optimization}
		Direct estimation of \eqref{eq:sum_rate_ul_mu_simo} requires the $L$ training sequences $\bm{b}_l$ 
		of the inter-cell interferers, so it is not a distributed method. 
		Therefore we introduce the \textit{distributed direct optimization} metric 
		$\max_{|\bm{w}|=\bm{1}}\sum_{k=1}^K \log_2 (1 + \gamma_k(\bm{w}))$, with
			\begin{multline} \label{eq:direct_terms}
				\gamma_k(\bm{w})=\\
				\frac{\left(\bm{b}_k\Hermitian\mathbf{P}_{\bm{Y}_{\bm{w}}\Hermitian}\bm{b}_k\right)^2/\sigma_k^2}
				{\sum_{\substack{k^{\prime}=1 \\ k^{\prime} \neq k}}^K 
				\left(\bm{b}_k\Hermitian\mathbf{P}_{\bm{Y}_{\bm{w}}\Hermitian}\bm{b}_{k^\prime}\right)^2/\sigma_{k^\prime}^2
				+T^2 \sigma_n^2 \bm{b}_k\Hermitian\left(\bm{Y}_{\bm{w}}\right)^+  \left(\bm{Y}_{\bm{w}}\Hermitian\right)^+\bm{b}_k} \text{,}
			\end{multline}
			where $\left(\bm{Y}_{\bm{w}}\right)^+$ is the Moore-Penrose Pseudo Inverse of $\bm{Y}_{\bm{w}}$.
			The LS filter $\bm{v}_k$ attempts to suppress both intra- and inter-cell interference,
			so that the optimization over $\bm{w}$, which depends on $\bm{v}_k$, implicitly takes
			the inter-cell interference directly into account. This scheme is suboptimal
			relative to the previous direct estimation scheme in which the BTS knows the training sequences
			for other-cell users, although the performance gap will be small given enough
			spatial degrees of freedom to suppress the inter-cell interference.

			Since LS filters converge to MMSE filters as $T \to \infty$, 
			the identity $\log_2\left(1+\mathrm{SINR}\right) = -\log_2(\mathrm{MMSE})=- \log_2 \left( \frac{1}{T}\norm{ \bm{v}_k\bm{Y}_{\bm{w}}-\bm{b}_k\Hermitian}^2_F\right)$ holds in the limit.
			Thus, we define the \textit{Least Squares Objective} by substituting the Least Squares Residual
			for the MMSE,
			\begin{equation}\label{eq:emp_mse_obj}
				\sum_{k=1}^K\log_2(\textrm{LS-Residual}_k) = \sum_{k=1}^K\log_2\left(\bm{b}_k\Hermitian\mathbf{P}^\perp_{\bm{Y}_{\bm{w}}\Hermitian}\bm{b}_k\right)\text{.}
			\end{equation}
			This cost function is somewhat simpler to work with than the direct estimate of SINR,
			although it suffers a relative performance degradation with limited training.
		
			Training for both DL channel estimation and direct IRS optimization 
			is performed in the UL direction, so that all UEs signals are received by each BTS.
			Then, minimizing the estimated objective gives the optimized IRS phase shifts $\bm{\Theta}^{(\mathrm{opt})}=\bm{w}^{(\mathrm{opt})}$ and receive filter $\bm{V}_{\bm{w}^{(\mathrm{opt})}}\Hermitian = [\bm{v}_1, \cdots, \bm{v}_K ]$.
			For both CSI-based schemes and direct optimization, as proposed here,
			the optimization over $\bm{\Theta}$ and $\bm{w}$ is non-convex and computationally complex.
			We will use a gradient type of algorithm,
			and leave further refinements that may reduce complexity for future work.

		\subsection{Forward-backward Training for IRS MIMO Channels}
			The MIMO extension of the direct IRS optimization method requires additional optimization 
			of the $K$ DL UE's receive filters, or scaled UL precoders, $\bm{G}$.
			For this we use synchronous bi-directional training with LS filters as proposed in \cite{Shi2014}.
			For the MC MIMO channel, as introduced in \eqref{eq:mu_mimo_system_model}, we first apply the proposed direct IRS optimization approach in the UL direction with an initial precoder $\bm{G}^0$.
			Then, the filters $\bm{V}$ and $\bm{G}$ are iteratively updated via bi-directional training.
			This method consists of two steps, the forward- and the backward update.
			In the forward update, all BTSs simultaneously send their DL training sequences 
			to their UEs, and apply the scaled LS filters estimated from UL training as precoders.
			Then, the UEs update their LS receive filters, based on the received DL training data.
			
			In the backward step, all UEs synchronously transmit their UL training sequences and 
			apply their normalized optimum receive filters from DL training as UL precoders.
			At the end of each iteration, the BTSs update their LS receive filters in parallel 
			based on their received UL training data and knowledge of intra-cell training sequences.

			Algorithm $1$
			shows the proposed direct optimization scheme for jointly optimizing the IRS phase shifts with precoders
			and combiners across multiple cells.
			In particular, we fix $\bm{\Theta}$ after the first initial backward update and joint optimization
			of $\bm{\Theta}$ and $\bm{V}$ in order to reduce training overhead.			
			The initial UL joint optimization requires $T(N_{\mathrm{IRS}}+1)$ training samples, 
			in contrast to a single-filter update, which requires $T$ training samples.
			The optimization procedure then requires $T(N_{\mathrm{IRS}}+1 + 2N_{\mathrm{FB}})$ training samples,
			where $N_{\mathrm{FB}}$ is the number of forward-backward iterations.
			Of course, this is suboptimal in that $\bm{\Theta}$ is optimized w.r.t. 
			the initial DL combiners $\bm{G}^0$.

			\begin{algorithm}
				\DontPrintSemicolon
				\TitleOfAlgo{Bi-directional Training for IRS-assisted MC MIMO Systems}
				Initialize all UE precoders $\bm{g}_k=\bm{1}$\;
				\tcp{IRS Optimization}
				All UEs synchronously transmit their training sequences $\frac{\bm{g}_k}{\norm{\bm{g}_k}_2}\overleftarrow{\bm{b}}_k\Hermitian$, $k=1,\cdots,K+L$, multiple ($N_{\textrm{IRS}}+1$) times with IRS phase shifts $\bm{\Theta}_j$, $j = 1,\dots,N_{\textrm{IRS}}+1$\;
				Each BTS jointly optimizes $\bm{V}$ and $\bm{w}$\;
				Set each cell's IRS to $\bm{\Theta}=\diag{\bm{w}}$\;
				\tcp{FB-Training}
				\For{$n=1$ \KwTo $N_{\mathrm{FB}}$}{
					\tcp{Forward Update}
					All BTS simultaneously send $\sum_{k=1}^K \frac{\bm{v}_k^*}{\norm{\bm{v}_k}_2}\overrightarrow{\bm{b}}\Hermitian_k$\;
					UEs update LS filters $\bm{g}_k =\left[\left(\overrightarrow{\bm{Y}} \overrightarrow{\bm{Y}}\Hermitian\right)^{-1} \overrightarrow{\bm{Y}}\overrightarrow{\bm{b}}_k\right]^*$\;
					\tcp{Backward Update}
					All UEs simultaneously send $\frac{\bm{g}_k}{\norm{\bm{g}_k}_2}\overleftarrow{\bm{b}}_k\Hermitian$\;
					BTS update LS filters $\bm{v}_k = \left(\overleftarrow{\bm{Y}} \overleftarrow{\bm{Y}}\Hermitian\right)^{-1}\overleftarrow{\bm{Y}}\overleftarrow{\bm{b}}_k$\;
				}
			\end{algorithm}

			For channel estimation only the UL training is needed. The IRS phase shifts and the filters
			can then be optimized via alternate updates (Max-SINR algorithm) offline without additional training.
			Similarly, the IRS phase shifts can be jointly updated with the UL receive filters.
			However, the BTSs have to exchange their cross-channel estimates
			since the UE's SINR terms for updating $\bm{g}_k$ in DL direction require CSI from each BTS to UE $k$.
			Also, in contrast to direct optimization, a channel estimation scheme must estimate 
			each entire $M_R \times N_T$ MIMO channel, whereas the proposed direct estimation schemes
			determine the filter (beamformer or combiner) coefficients, which will be a much smaller
			set in the MC scenario with many antennas per node.

			\section{Simulation Results} \label{sec:simulation_results}
			We next present a set of numerical results that compare the sum-rate performance
			of the proposed direct optimization schemes with channel estimation. We first
			describe the set of methods in the comparison.
			\subsection{Schemes Compared}
				\subsubsection{Perfect CSI}
					 jointly maximizes \eqref{eq:sum_rate_ul_mu_simo} w.r.t. 
					$\bm{\Theta}$ and $\bm{v}_k$'s assuming perfect knowledge of all channels.
					This serves as an ideal benchmark.

				\subsubsection{Full Channel Estimation} estimates UL channels of {\em all} 
				users (intra- and inter-cell).  This serves as a centralized benchmark
				where all cross-channels are available. For MISO channels the training data 
				is the same as for the proposed direct IRS optimization scheme.

				\subsubsection{Partial Channel Estimation} estimates only the channels for the $K$
				intra-cell users. The IRS parameters and the UL receive filters are then optimized 
				for the objective \eqref{eq:sum_rate_ul_mu_simo}, which neglects the $L$ interferers.
				The channel estimation and decomposition scheme from \cite{Zhou2021a} is used
				to estimate the direct and the composite IRS channels.

				\subsubsection{Direct Centralized Optimization} extends the objective \eqref{eq:direct_terms} 
				by including all $L$ interference terms. This is then analogous to Full Channel Estimation.
				
				\subsubsection{Direct Decentralized Optimization} uses the objective \eqref{eq:direct_terms}.
				In addition, we also present results for the \textit{LS Objective} in \eqref{eq:emp_mse_obj}.
				This is then analogous to Partial Channel Estimation.

				\subsubsection{Direct Filter Estimation with Random $\bm{\Theta}$} assigns 
				$\bm{\Theta}$ iid uniform phase shifts, which are fixed. 
				The filters $\bm{v}_k$ and $\bm{g}_k$ are optimized with decentralized bi-directional training
				according to Algorithm $1$ with fixed $\bm{\Theta}$. This serves as a benchmark
				to assess the gains offered by optimizing the IRS parameters.

	\subsection{System Parameters} \label{sec:twocel}
		We consider two cells each with $K=2$ UEs.
		Each cell has a single IRS with $N_{\mathrm{IRS}}=16$ elements,
		the BTS has $M_R=6$ antennas, and the UE's have a single antenna
		for the MISO scenarios, and $N_\mathrm{T}=2$ antennas for the MIMO scenarios.
		The direct channels $\bm{H}_{\mathrm{RT}}$ and the UE-to-IRS channels $\bm{H}_{\mathrm{IT}}$ 
		are full-scattering, i.e., their elements are iid. zero-mean complex Gaussian.
		The intra-cell channels are unit-variance and the cross-channels are $3\mathrm{dB}$ weaker.
		The BTS-to-IRS channel $\bm{H}_{\mathrm{IR}}\Transpose$ follows a 
		Rician channel model, as in \cite{Wu2019a}, with a Rician factor $\beta_R=10$, 
		which corresponds to a strong LoS component.
		The elements of $\bm{H}_{\mathrm{IR}}\Transpose$ also have $\mu_h^2+\sigma_h^2=1$, transmit power $\sigma_m^2=0\mathrm{dB}$ and noise variance
		$\sigma_n^2=10\mathrm{dB}$ where unspecified.

		A gradient ascent method with an initial Greedy search and Armijo step size rule was used
		to optimize the IRS phases $\varphi_j$ where $w_j=\e^{j\varphi_j}$.
		The optimization is for the UL sum rate with linear receivers, but the results are shown for the DL objective
		since the optimized UL combiners are used for the DL beamformers.
		The results show an average over $100$ Monte Carlo runs.
		
		\subsection{MC MISO Channels}
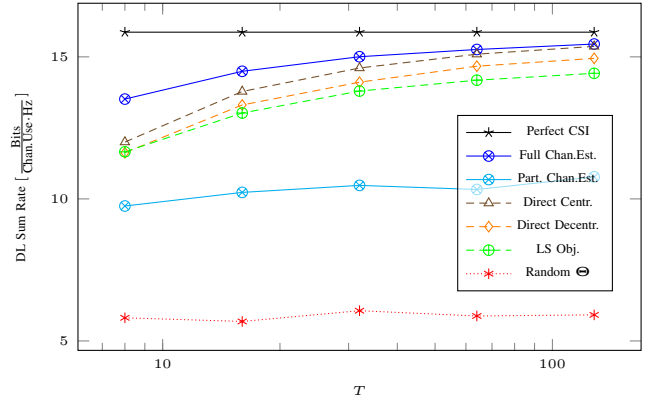
\begin{figure}
				\centering
				\begin{tikzpicture}
					\begin{semilogxaxis}[
					cycle list name=mydefaultlist,
					every axis/.append style={font=\tiny},
					ylabel={DL Sum Rate $[\frac{\textrm{Bits}}{\textrm{Chan.Use} \cdot \textrm{Hz}}]$},
					xlabel={$T$},
					width=0.5\textwidth,
					height=6.2cm,
					xtick={1, 10, 100},
					xticklabels={$1$, $10$,$100$},
					legend style={at={(0.95,0.17)}, 
					anchor=south east, legend columns=1, fill=white, fill opacity=0.6, draw opacity=1, text opacity =1}
					]
						\addplot table [x=N_samples, y=Perfect_CSI, col sep=comma]{./data/mc_miso/mc_miso_chan_est_DLSR_sigmaNsq10dB69.csv};
						\addlegendentry{Perfect CSI}
						\addplot table [x=N_samples, y=Channel_Est, col sep=comma]{./data/mc_miso/mc_miso_chan_est_DLSR_sigmaNsq10dB69.csv};
						\addlegendentry{Full Chan.Est.}
						\addplot table [x=N_samples, y=Channel_Est-NoIntf, col sep=comma]{./data/mc_miso/mc_miso_chan_est_DLSR_sigmaNsq10dB69.csv};
						\addlegendentry{Part. Chan.Est.}
						\addplot table [x=N_samples, y=PreciseObj, col sep=comma]{./data/mc_miso/mc_miso_chan_est_DLSR_sigmaNsq10dB69.csv};
						\addlegendentry{Direct Centr.}
						\addplot table [x=N_samples, y=NoIntf, col sep=comma]{./data/mc_miso/mc_miso_chan_est_DLSR_sigmaNsq10dB69.csv};
						\addlegendentry{Direct Decentr.}
						\addplot table [x=N_samples, y=AugmentedSimplifieda1, col sep=comma]{./data/mc_miso/mc_miso_chan_est_DLSR_sigmaNsq10dB69.csv};
						\addlegendentry{LS Obj.}
						\addplot table [x=N_samples, y=RandomTheta, col sep=comma]{./data/mc_miso/mc_miso_chan_est_DLSR_sigmaNsq10dB69.csv};
						\addlegendentry{Random $\bm{\Theta}$}
					\end{semilogxaxis}
				\end{tikzpicture}
				\caption{DL Sum Rate vs. $T$ for MISO channels.}
				\label{fig:mc_dl_miso_6x1}
			\end{figure}

			Figure \ref{fig:mc_dl_miso_6x1} shows the DL sum rate over all users versus number
			of training symbols for MISO channels. For all plots in this section the solid curves 
			correspond to channel estimation schemes and the dashed curves correspond
			to direct estimation. The results show a modest performance degradation for the direct
			estimation methods compared with full channel estimation, where the gap narrows
			as $T$ increases. Using the LS objective does not perform as well as direct estimation of
			the SINR since with small $T$, the LS objective does not accurately approximate the MMSE.
			The distributed schemes perform much better than partial channel estimation since
			the inter-cell interference is significant in this scenario.

\begin{figure}
				\centering
				\begin{tikzpicture}
					\begin{semilogxaxis}[
					cycle list name=mydefaultlist,
					every axis/.append style={font=\tiny},
					ylabel={DL Sum Rate $[\frac{\textrm{Bits}}{\textrm{Chan.Use} \cdot \textrm{Hz}}]$},
					xlabel={$-10\log(\sigma_n^2) \text{ } [\mathrm{dB}]$},
					width=0.5\textwidth,
					height=6.2cm,
					xtick={0.001,0.01,0.1,1},
					xticklabels={$-30$,$-20$,$-10$,$0$},
					legend style={at={(0.05,0.95)}, 
					anchor=north west, legend columns=1, fill=white, fill opacity=0.6, draw opacity=1, text opacity =1}
					]
						\addplot table [x=one_over_sigma_n_sq, y=Perfect_CSI, col sep=comma]{./data/mc_miso/mc_miso_chan_est_DLSR_nsamples16-992.csv};
						\addlegendentry{Perfect CSI}
						\addplot table [x=one_over_sigma_n_sq, y=Channel_Est, col sep=comma]{./data/mc_miso/mc_miso_chan_est_DLSR_nsamples16-992.csv};
						\addlegendentry{Full Chan.Est.}
						\addplot table [x=one_over_sigma_n_sq, y=Channel_Est-NoIntf, col sep=comma]{./data/mc_miso/mc_miso_chan_est_DLSR_nsamples16-992.csv};
						\addlegendentry{Part. Chan.Est.}
						\addplot table [x=one_over_sigma_n_sq, y=PreciseObj, col sep=comma]{./data/mc_miso/mc_miso_chan_est_DLSR_nsamples16-992.csv};
						\addlegendentry{Direct Centr.}
						\addplot table [x=one_over_sigma_n_sq, y=NoIntf, col sep=comma]{./data/mc_miso/mc_miso_chan_est_DLSR_nsamples16-992.csv};
						\addlegendentry{Direct Decentr.}
						\addplot table [x=one_over_sigma_n_sq, y=AugmentedSimplifieda1, col sep=comma]{./data/mc_miso/mc_miso_chan_est_DLSR_nsamples16-992.csv};
						\addlegendentry{LS Obj.}
						\addplot table [x=one_over_sigma_n_sq, y=RandomTheta, col sep=comma]{./data/mc_miso/mc_miso_chan_est_DLSR_nsamples16-992.csv};
						\addlegendentry{Random $\bm{\Theta}$}
					\end{semilogxaxis}
				\end{tikzpicture}
				\caption{DL Sum Rate vs. $1/\sigma_n^2$ (dB) with MISO channels.}
				\label{fig:mc_dl_miso_6x1_sr_vs_sigma_n_sq}
			\end{figure}
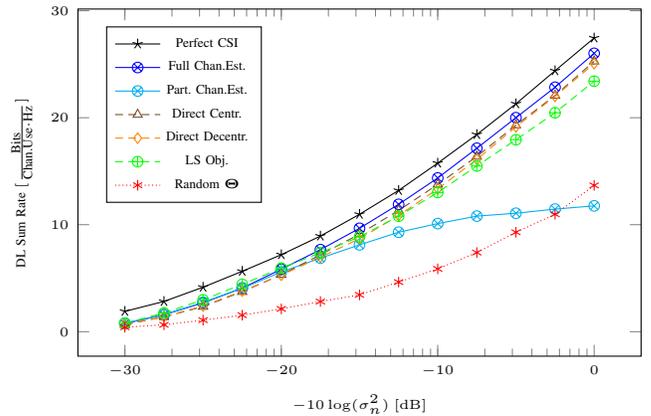

			Figure \ref{fig:mc_dl_miso_6x1_sr_vs_sigma_n_sq} shows the DL sum rate versus $1/\sigma_n^2$ with $T=16$.
			The direct optimization schemes perform close to full channel estimation.
			Partial channel estimation shows a saturation effect as the interference becomes dominant relative to noise. 

		\subsection{MC MIMO Channels}
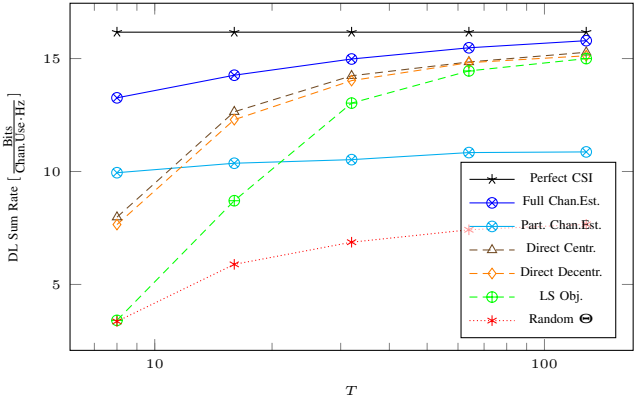
\begin{figure}
				\centering
				\begin{tikzpicture}
					\begin{semilogxaxis}[
					cycle list name=mydefaultlist,
					every axis/.append style={font=\tiny},
					ylabel={DL Sum Rate $[\frac{\textrm{Bits}}{\textrm{Chan.Use} \cdot \textrm{Hz}}]$},
					xlabel={$T$},
					width=0.5\textwidth,
					height=6.2cm,
					xtick={1, 10, 100},
					xticklabels={$1$, $10$,$100$},
					legend style={at={(0.97,0.035)}, 
					anchor=south east, legend columns=1, fill=white, fill opacity=0.6, draw opacity=1, text opacity =1}
					]
						\addplot table [x=N_samples, y=Perfect_CSI, col sep=comma]{./data/mc_mimo/mc_mimo_chan_est_DLSR_sigmaNsq10dB629.csv};
						\addlegendentry{Perfect CSI}
						\addplot table [x=N_samples, y=Channel_Est, col sep=comma]{./data/mc_mimo/mc_mimo_chan_est_DLSR_sigmaNsq10dB629.csv};
						\addlegendentry{Full Chan.Est.}
						\addplot table [x=N_samples, y=Channel_Est-NoIntf, col sep=comma]{./data/mc_mimo/mc_mimo_chan_est_DLSR_sigmaNsq10dB629.csv};
						\addlegendentry{Part. Chan.Est.}
						\addplot table [x=N_samples, y=PreciseObj, col sep=comma]{./data/mc_mimo/mc_mimo_chan_est_DLSR_sigmaNsq10dB629.csv};
						\addlegendentry{Direct Centr.}
						\addplot table [x=N_samples, y=NoIntf, col sep=comma]{./data/mc_mimo/mc_mimo_chan_est_DLSR_sigmaNsq10dB629.csv};
						\addlegendentry{Direct Decentr.}
						\addplot table [x=N_samples, y=LogLSObj, col sep=comma]{./data/mc_mimo/mc_mimo_chan_est_DLSR_sigmaNsq10dB629.csv};
						\addlegendentry{LS Obj.}
						\addplot table [x=N_samples, y=RandomTheta, col sep=comma]{./data/mc_mimo/mc_mimo_chan_est_DLSR_sigmaNsq10dB629.csv};
						\addlegendentry{Random $\bm{\Theta}$}
					\end{semilogxaxis}
				\end{tikzpicture}
				\caption{DL Sum Rate vs. $T$ for MIMO channels.}
				\label{fig:mc_dl_mimo_6x2}
			\end{figure}

			Figure \ref{fig:mc_dl_mimo_6x2} shows the DL sum rate versus training symbols $T$ with MIMO channels.
			The CSI-based schemes performed three iterations for the offline alternating optimization to compute
			the filters and IRS parameters, which was observed	to be adequate for convergence.
			The proposed direct optimization scheme performed $N_{\mathrm{FB}}=2$ iterations in Algorithm $1$.
			
			The trends are similar to those shown for the MISO case,
			although the performance loss for direct optimization is more pronounced
			when $T$ is small, especially for the LS objective.
			For $T>16$ distributed optimization achieves significantly better performance than partial channel estimation.
		\subsection{Performance Comparison When Increasing $N_{\mathrm{IRS}}$}
\begin{figure}
				\centering
				\begin{tikzpicture}
					\begin{semilogxaxis}[
					cycle list name=mydefaultlist,
					every axis/.append style={font=\tiny},
					ylabel={UL Rate $[\frac{\textrm{Bits}}{\textrm{Chan.Use} \cdot \textrm{Hz}}]$},
					xlabel={$N_{\mathrm{IRS}}$},
					width=0.5\textwidth,
					height=6.2cm,
					ymin=0,
					xtick={1, 10, 100},
					xticklabels={$1$, $10$,$100$},
					legend style={at={(0.05,0.95)}, 
					anchor=north west, legend columns=1, fill=white, fill opacity=0.6, draw opacity=1, text opacity =1}
					]
						\addplot table [x=N_IRS, y=Perfect_CSI, col sep=comma]{./data/increasing_N_IRS/mu_miso_K_vs_N_IRS_SRvsK_sigma_N_sq10dB506_1.csv};
						\addlegendentry{Perfect CSI}
						\addplot table [x=N_IRS, y=Channel_Est, col sep=comma]{./data/increasing_N_IRS/mu_miso_K_vs_N_IRS_SRvsK_sigma_N_sq10dB506_1.csv};
						\addlegendentry{Full Chan.Est.}
						\addplot table [x=N_IRS, y=Channel_Est-NoIntf, col sep=comma]{./data/increasing_N_IRS/mu_miso_K_vs_N_IRS_SRvsK_sigma_N_sq10dB506_1.csv};
						\addlegendentry{Part. Chan.Est.}
						\addplot table [x=N_IRS, y=PreciseObj, col sep=comma]{./data/increasing_N_IRS/mu_miso_K_vs_N_IRS_SRvsK_sigma_N_sq10dB506_1.csv};
						\addlegendentry{Direct Centr.}
						\addplot table [x=N_IRS, y=NoIntf, col sep=comma]{./data/increasing_N_IRS/mu_miso_K_vs_N_IRS_SRvsK_sigma_N_sq10dB506_1.csv};
						\addlegendentry{Direct Decentr.}
						\addplot table [x=N_IRS, y=LogLSObj, col sep=comma]{./data/increasing_N_IRS/mu_miso_K_vs_N_IRS_SRvsK_sigma_N_sq10dB506_1.csv};
						\addlegendentry{LS Obj.}
						\addplot table [x=N_IRS, y=RandomTheta, col sep=comma]{./data/increasing_N_IRS/mu_miso_K_vs_N_IRS_SRvsK_sigma_N_sq10dB506_1.csv};
						\addlegendentry{Random $\bm{\Theta}$}
					\end{semilogxaxis}
				\end{tikzpicture}
				\caption{UL single-user rate vs. $N_{\mathrm{IRS}}$ for MISO channels.}
				\label{fig:growing_N_IRS}
			\end{figure}
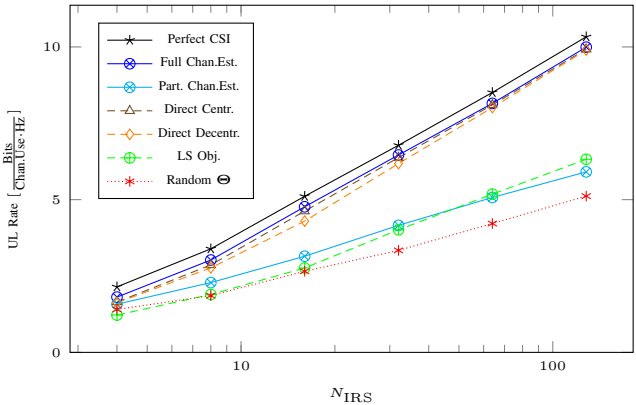

			Figure \ref{fig:growing_N_IRS} shows plots of the UL single-user rate vs. the number 
			of IRS elements $N_{\mathrm{IRS}}$ for MISO channels.
			Here $K=1$ and $L=3$ in \eqref{eq:mu_simo_system_model}, and
			the number of training samples per epoch (fixed $\bm{\Theta}_j$) is $T=16$.
			Also, for these results all channels including $\bm{H}_{\mathrm{IR}}\Transpose$ are full scattering.
			Distributed SINR optimization only has a small performance loss compared to
			full channel estimation, where the gap diminishes for larger values of $N_{\mathrm{IRS}}$.
			As in the previous simulations, the LS objective suffers a significant performance loss
			due to the low number of training symbols although
			for larger values of $N_{\mathrm{IRS}}$, the performance improves 
			relative to random IRS phase-shifts.
			Partial channel estimation performs relatively poorly since it ignores
			the interference.
		
		\subsection{Further Observations}
			In scenarios with strong LoS paths or rank-one channels it is not possible 
			to distinguish the MIMO channels, which results in increased interference.
			For larger numbers of BTS antennas $M_{\mathrm{R}}$, the performance gap 
			between our proposed method and partial channel estimation increases
			since spatial beamforming allows better separation of UEs and interferers. 

			Finally, we remark that optimizing the phase-shifts of $\bm{w}$ is highly non-convex.
			By using a gradient-descent algorithm, we sometimes encountered numerical issues 
			with convergence to local optima, especially with few degrees of freedom 
			(low $N_{\mathrm{IRS}}$, $M_{\mathrm{R}}$), at higher SNR and large $T$.
			This is similar to other optimization methods for IRS phase-shifts, based on channel estimates,
			such as Semi-Definite Relaxation with Gaussian randomization methods, 
			as applied in \cite{Yang2020}, which also may not find the global optimum.
			We leave the investigation of more efficient algorithms for direct optimization
			of IRS parameters and precoders for future work. 
			
	\section{Conclusions} \label{sec:conclusion}
		We have presented techniques for jointly optimizing IRS phase-shifts with precoders across multiple
		cells that do not rely on explicit channel estimation. In the distributed versions, each BTS knows
		the training sequences for only a subset of UEs, e.g., for those it serves. This eliminates the need
		for coordination between cells beyond synchronized training. For the scenarios considered the distributed method 
		effectively suppresses inter-cell interference and
		shows only a modest performance degradation compared with estimating
		all MISO or MIMO channels across cells.
		The scheme is scalable in the sense that the training overhead
		scales linearly with the number of IRS elements, but does not need to scale
		with the number of users or cells. Future work consists of combining the proposed schemes
		with power control and allowing for multiple transmitted streams.

	\bibliographystyle{IEEEtran}
	\bibliography{IEEEabrv,bibliography}

\end{document}